\documentclass[twocolumn]{aastex62}

\received{}
\revised{}
\accepted{}

\submitjournal{The Astrophysical Journal}

\usepackage{gensymb}

\shorttitle{Ion Injection in ICM Shocks}
\shortauthors{Ha et al.}

\begin{document}

\title{Proton Acceleration in Weak Quasi-parallel Intracluster Shocks: Injection and Early Acceleration}

\author[0000-0001-7670-4897]{Ji-Hoon Ha}
\affiliation{Department of Physics, School of Natural Sciences, UNIST, Ulsan 44919, Korea}
\author[0000-0002-5455-2957]{Dongsu Ryu}
\affiliation{Department of Physics, School of Natural Sciences, UNIST, Ulsan 44919, Korea}
\author[0000-0002-4674-5687]{Hyesung Kang}
\affiliation{Department of Earth Sciences, Pusan National University, Busan 46241, Korea}
\author{Allard Jan van Marle}
\affiliation{Department of Physics, School of Natural Sciences, UNIST, Ulsan 44919, Korea}
\correspondingauthor{Dongsu Ryu}
\email{ryu@sirius.unist.ac.kr}

\begin{abstract}


Collisionless shocks with low sonic Mach numbers, $M_{\rm s} \lesssim 4$, are expected to accelerate cosmic ray (CR) protons via diffusive shock acceleration (DSA) in the intracluster medium (ICM). However, observational evidence for CR protons in the ICM has yet to be established.
Performing particle-in-cell simulations, we study the injection of protons into DSA and the early development of a nonthermal particle population in weak shocks in high $\beta$ ($\approx 100$) plasmas.
Reflection of incident protons, self-excitation of plasma waves via CR-driven instabilities, and multiple cycles of shock drift acceleration are essential to the early acceleration of CR protons in supercritical quasi-parallel shocks.
We find that only in ICM shocks with $M_{\rm s} \gtrsim M_{\rm s}^*\approx 2.25$, a sufficient fraction of incoming protons are reflected by the overshoot in the shock electric potential and magnetic mirror at locally perpendicular magnetic fields,
leading to efficient excitation of magnetic waves via CR streaming instabilities and the injection into the DSA process. 
Since a significant fraction of ICM shocks have $M_{\rm s} < M_{\rm s}^*$, CR proton acceleration in the ICM might be less efficient than previously expected.
This may explain why the diffuse gamma-ray emission from galaxy clusters due to proton-proton collisions has not been detected so far.

\end{abstract}


\section{Introduction}
\label{sec:s1}

Hierarchical clustering of the large-scale structure of the universe induces supersonic flow motions of baryonic matter, which result in the formation of weak shocks with sonic Mach numbers $M_{\rm s} \lesssim 4$ in the hot intracluster medium (ICM) \citep[e.g.,][]{miniati2000, ryu2003}. 
The properties of these structure formation shocks and the energy dissipation in the shocks have been extensively studied through cosmological hydrodynamic simulations \citep[e.g.,][]{miniati2000, ryu2003, pfrommer2006, kang2007, skillman2008, hoeft2008, vazza2009, hong2014, schaal2015, hong2015}.
In particular, shocks associated with mergers of sub-cluster clumps have been observed in X-ray and radio \citep[e.g.,][]{markevitch07,vanweeren10,bruggen2012,brunetti2014}, and also studied by simulations \cite[e.g.,][]{paul2011,schmidt2017,ha2018}.

Just like Earth's bow shocks and supernova remnant shocks, ICM shocks are thought to accelerate cosmic ray (CR) protons and electrons via diffusive shock acceleration (DSA, {\it a.k.a.} Fermi I acceleration) \citep[e.g.,][]{bell1978, blandford1978, drury1983}.
Although the acceleration of relativistic electrons can be inferred from the so-called giant radio relics such as the Sausage relic in the merging cluster CIZA J2242.8+5301 \citep{vanweeren10}, the presence of the CR protons produced by ICM shocks has yet to be confirmed \citep[e.g.,][]{pinzke2010,zandanel2014,kang2018}.
Inelastic collisions of CR protons with thermal protons followed by the decay of neutral pions produce diffuse gamma-ray emission, which has not been detected so far with {\it Fermi}-LAT \citep{ackermann2016}.
According to studies using cosmological hydrodynamic simulations that adopt prescriptions for the {\it CR proton acceleration efficiency} in shocks, $\eta( M_{\rm s})$,
the non-detection of gamma-ray emission from galaxy clusters constrains $\eta(M)$ to be less than $10^{-3}$ for $2\le M_{\rm s}\le5$ \citep[e.g.,][]{vazza2016}.

Collisionless astrophysical shocks involve complex kinetic plasma processes, such as wave-generations and wave-particle interactions, well beyond those described by the magnetohydrodynamics (MHD) Rankine-Hugoniot jump condition in collisional shocks \citep[see, e.g.,][for a review]{treumann2009}.
The key element in estimating the DSA acceleration efficiency $\eta$ is the so-called `injection process', which energizes thermal protons to the suprathermal energies sufficient to diffuse across the shock.
In efforts to understand CR injection and early acceleration, space/astrophysicists have investigated kinetic processes around shocks using particle-in-cell (PIC) and hybrid plasma simulations \citep[e.g.,][]{guo2014a, guo2014b,caprioli2014a, caprioli2014b,caprioli2015, park2015} \citep[see also][and references therein]{treumann2009}. 
In PIC simulations, both ions and electrons are treated kinetically, and therefore various microinstabilities and wave-particle interactions can be followed from first principles. 
In hybrid simulations, on the other hand, only ions are treated kinetically, while electrons are modeled to be charge-neutralizing fluids with zero-mass.
They are suitable for studying not only ion injection but also long-term acceleration during the Fermi I regime, since they are computationally much less expensive than PIC simulations.

Comprehensive studies using hybrid simulations showed that CR ions are efficiently accelerated with $\eta\sim 0.05-0.15$ in strong, `quasi-parallel shocks' with $M_{\rm s}\gtrsim 5$ and $\theta_{\rm Bn} \lesssim 45^{\circ}$ \citep{caprioli2014a,caprioli2014b,caprioli2015}. Here, $\theta_{\rm Bn}$ is the obliquity angle between the shock normal and the background magnetic field direction. It is one of the key parameters that govern the characteristics of shocks; quasi-parallel shocks have $\theta_{\rm Bn}\lesssim 45^{\circ}$, while quasi-perpendicular shocks have $\theta_{\rm Bn}\gtrsim 45^{\circ}$.
In particular, \citet{caprioli2015} presented that in quasi-parallel shocks a substantial fraction of ions impinging on the shock potential barrier
can be specularly reflected, when the quasi-periodically reforming shock potential, $\phi$, is in a high state (i.e., $e\Delta \phi > m_i v_x^2 / 2$).
The reflected ions escaping upstream along parallel magnetic fields generate low-frequency waves and amplify transverse magnetic fields via CR ion-driven instabilities, transforming a part of upstream quasi-parallel fields to locally quasi-perpendicular fields in the shock transition layer.
Then, ions arriving subsequently at the shock can be reflected at the locally perpendicular portions of turbulent magnetic fields \citep[see also][]{sundberg2016}.
With the transverse magnetic fields, the reflected ions gain sufficient energies via multiple cycles of shock drift acceleration (SDA), and then start participating in the Fermi I cycle of shock acceleration.
Hence, reflection of ions, self-excitation of turbulent waves, and SDA are the integral parts of ion injection and acceleration in quasi-parallel shocks.

In `quasi-perpendicular shocks', on the other hand, although ions can be reflected by the magnetic mirror force due to converged magnetic field lines at the shock ramp, they are expected to advect downstream along with the background fields typically after one gyromotion. According to hybrid simulations, reflected ions may undergo only a few cycles of SDA, but do not reach the energies sufficient to be injected to Fermi I process \citep{caprioli2014a}. Since the gyrostream of reflected ions penetrates upstream less than one ion gyroradius from the shock ramp, turbulent waves are not efficiently excited in the precursor of quasi-perpendicular shocks \citep{caprioli2014b}.
However, recent simulations using an approach combining PIC and MHD codes showed that if ions are `injected' with a sufficient amount, they could be further energized by SDA and excite the shock corrugation instability \citep{vanmarle2018}. The DSA of ions in the quasi-perpendicular configuration, hence, needs to be further investigated.

The criticality for particle reflection at collisionless shocks has been studied for Earth's bow shocks and interplanetary shocks. \citet{edmiston1984} calculated the `fast first critical Mach number', $M_{\rm f}^*(\beta,\theta_{\rm Bn})$ for $0\le \beta \le 4$ and $0^{\circ} \le \theta_{\rm Bn} \le 90^{\circ}$, from the condition that the downstream flow speed normal to the shock equals the downstream sound speed, $v_{\rm n2}=c_{\rm s2}$. Here, $\beta \equiv P_{\rm gas}/P_{\rm B}$ is the plasma beta. In `supercritical' shocks with fast Mach number $M_{\rm f} \gtrsim M_{\rm f}^*$, the shock energy cannot be dissipated through resistivity alone, so a substantial fraction of incoming ions must be reflected upstream and/or dispersive waves with sufficient energy fluxes must be emitted upstream in order to satisfy the MHD Rankine-Hunoniot jump. The critical Mach number $M_{\rm f}^* \approx 2.76$ is frequently quoted\footnote{The Alfv\'en Mach number $M_{\rm A}$ is often used to characterize collisionless shocks with $\beta\sim 1$, and hence, the critical Mach number is often given as $M_{\rm A}^* \approx 2.76$.}, but it is for shocks with $\beta = 0$ and $\theta_{\rm Bn} = 90^{\circ}$,
that is, for perpendicular shocks with strong background magnetic fields. The critical Mach number for quasi-parallel shocks in typical astrophysical environments with $\beta\sim 1$, for instance, is estimated to be $M_{\rm f}^* \approx 1.0-1.5$. It becomes smaller for higher $\beta$ and for smaller $\theta_{\rm Bn}$.

Turbulent waves, which take part in the injection and acceleration of CR protons, are induced by two dominant modes: (1) resonant streaming instability which excites left-handed circularly polarized waves \citep{bell1978}, and
(2) nonresonant current-driven instability which excites right-handed circularly polarized waves \citep{bell2004}.
Using hybrid simulations of quasi-parallel shocks in $\beta\sim 1$ plasmas, \citet{caprioli2014b} argued that resonant streaming instability is dominant in the precursor of shocks with Alfv\'en Mach number $M_{\rm A} \lesssim 30$, while nonresonant current-driven instability operates faster in stronger shocks with $M_{\rm A} \gtrsim 30$.
Both instabilities amplify primarily the transverse component of magnetic fields, so they generate locally perpendicular fields in the foreshock and downstream regions, which in turn reflect subsequently arriving ions and facilitate the SDA of reflected ions.
The increase in the magnetic energy due to these instabilities scales linearly with $M_{\rm A}$, i.e., $(\delta B/B_0)^2 \propto M_{\rm A}$.
Eventually, excited turbulent waves act as scattering centers both upstream and downstream of the shock, which are required for DSA.

The above cited papers by Caprioli and collaborators investigated proton acceleration in strong shocks with high $M_{\rm A}$ ($\ge 5$) for mostly $\beta \sim 1$ plasmas, which are expected to be all supercritical.
In hot ICM plasmas, where $\beta \sim 100$ \citep[e.g.,][]{ryu2008,porter2015}, shocks have low sonic Mach numbers, $M_{\rm s} \approx 2-4$, but relatively high Alfv\'en Mach numbers, $M_{\rm A} \approx 10 M_{\rm s} \approx 20-40$, which are much higher than $M_{\rm A}^*=2.76$. 
However, we presume that in high $\beta$ regimes, instead of $M_{\rm A}$, the sonic Mach number $M_s$, which determines the shock compression ratio, is the more relevant parameter in defining the shock criticality,
since ion reflection is governed by the overshoot, $\Delta \phi$, and the magnetic field mirror force due to compressed magnetic field lines at the shock ramp.
Moreover, in high $\beta$ plasmas with high temperature and weak magnetic field strength, ion reflection may be suppressed at lower $M_{\rm s}$ due to the smoothing of $\phi$ by fast thermal motions and weaker magnetic mirror force.
To our knowledge, the injection and DSA of CR protons in weak, `quasi-parallel shocks' in such high $\beta$ environments have not yet been investigated with simulations.
Note that for these ICM shocks, $M_{\rm f} \approx M_{\rm s}$.

\citet{kraussvarban1991} studied quasi-parallel shocks with low Alfv\'en Mach numbers, $M_A<3.5$, in $\beta\approx 1$ plasmas with hybrid simulations.
They found that at $M_{\rm A}\lesssim 1.5$, shocks are steady and subcritical with little back-streaming ions, and phase-standing whistlers are dominant in the foreshock region.
At $M_{\rm A}\gtrsim 2.3$, on the other hand, shocks become unsteady and undergo cyclic self-reformation due to the accumulation of reflected ions, which excite fast magnetosonic dispersive whistlers with wavelengths longer than phase-standing whistlers.
So the transition from low Mach, steady, subcritical shocks to high Mach, self-reforming, supercritcal shocks seems to occur around $M_{\rm A}\approx M_{\rm s}\sim 2.3$ in $\beta \approx 1$ environments, and it is closely related with the reflection of incoming ions at the shock ramp.

Electron acceleration in `quasi-perpendicular shocks' in high $\beta$ ICM plasmas was studied using PIC simulations before \citep{guo2014a, guo2014b}.
It was shown that in shocks with $M_{\rm s}=3$, $\beta=20$, and $\theta_{\rm Bn}=63^{\circ}$, for instance, about 20 \% of incoming ions are reflected and gain a small amount of energy via a few cycles of SDA. However, those ions pass through the potential barrier and advect downstream along with the background magnetic field.
Besides, the simulations by \citet{guo2014a, guo2014b} did not extend to many ion gyration periods, which are necessary for studies of ion acceleration, since their primary focus was on electron acceleration.

In this paper, we examine the physics of `shock criticality' and the injection and early acceleration of CR protons in weak, quasi-parallel shocks in high $\beta$ ICM plasmas.
Considering that the early development of collisionless shock formation involves kinetic processes due to both electrons and protons, PIC simulations are employed; but then simulations are limited to be either one-dimensional (1D) or two-dimensional (2D) in less than a hundred ion gyration periods.
We inspect the injection and early acceleration of ions, along with shock structures and ion energy spectra. The nature of CR ion-driven instabilities and turbulent magnetic field amplification is probed with Fourier analyses of upstream self-excited magnetic fields. We also discuss the dependence of ion injection and CR ion-driven instabilities on the pre-shock conditions, such as $M_{\rm s}$, $\beta$, and $\theta_{\rm Bn}$.

This paper is organized as follows. In Section \ref{sec:s2}, numerical details of PIC simulations are presented. 
In Section \ref{sec:s3}, ion injection and CR ion-driven instabilities are described and the dependence on various shock parameters are discussed. 
A brief summary follows in Section \ref{sec:s4}. 

\section{Numerics}
\label{sec:s2}

\begin{deluxetable*}{ccccccccccccc}[t]
\tablecaption{Model Parameters for Simulations$^a$ \label{tab:t1}}
\tabletypesize{\scriptsize}
\tablecolumns{13}
\tablenum{1}
\tablewidth{0pt}
\tablehead{
\colhead{Model Name} &
\colhead{$M_{\rm s}\approx M_{\rm f} $} &
\colhead{$M_{\rm A}$} &
\colhead{$v_0/c$} &
\colhead{$\theta_{\rm Bn}$} &
\colhead{$\beta$} &
\colhead{$T_e = T_i [\rm K(keV)]$} &
\colhead{$m_i / m_e$} &
\colhead{$L_x [c/w_{\rm pe}]$} &
\colhead{$L_y [c/w_{\rm pe}]$} &
\colhead{$\Delta x [c/w_{\rm pe}]$} &
\colhead{$t_{\rm end} [w_{\rm pe}^{-1}]$}&
\colhead{$t_{\rm end} [\Omega_{\rm ci}^{-1}]^c$}
}
\startdata
M3.2$^b$ & 3.2 & 29.2 & 0.052 & $13^{\circ}$ &100 & $10^8(8.6)$ & 100 & $2\times 10^4$ & 2 & 0.1 & $3.4\times 10^5$ & 90.2\\
\hline
M2.0 & 2.0 & 18.2 & 0.027 & $13^{\circ}$ &100 & $10^8(8.6)$ & 100 & $2\times 10^4$ & 2 & 0.1 & $3.4\times 10^5$ & 90.2\\
M2.15 & 2.15 & 19.6 & 0.0297 & $13^{\circ}$ &100 & $10^8(8.6)$ & 100 & $2\times 10^4$ & 2 & 0.1 & $3.4\times 10^5$ & 90.2\\
M2.25 & 2.25 & 20.5 & 0.0315 & $13^{\circ}$ &100 & $10^8(8.6)$ & 100 & $2\times 10^4$ & 2 & 0.1 & $3.4\times 10^5$ & 90.2\\
M2.5 & 2.5 & 22.9 & 0.035 & $13^{\circ}$ &100 & $10^8(8.6)$ & 100 & $2\times 10^4$ & 2 & 0.1 & $3.4\times 10^5$ & 90.2\\
M2.85 & 2.85 & 26.0 & 0.0395 & $13^{\circ}$ &100 & $10^8(8.6)$ & 100 & $2\times 10^4$ & 2 & 0.1 & $3.4\times 10^5$ & 90.2\\
M3.5 & 3.5 & 31.9 & 0.057 & $13^{\circ}$ &100 & $10^8(8.6)$ & 100 & $2\times 10^4$ & 2 & 0.1 & $3.4\times 10^5$ & 90.2\\
M4 & 4.0 & 36.5 & 0.066 & $13^{\circ}$ &100 & $10^8(8.6)$ & 100 & $2\times 10^4$ & 2 & 0.1 & $3.4\times 10^5$ & 90.2\\
\hline
M3.2-$\theta$23 & 3.2 & 29.2 & 0.052 & $23^{\circ}$ &100 & $10^8(8.6)$ & 100 & $2\times 10^4$ & 2 & 0.1 & $3.4\times 10^5$ & 90.2\\
M3.2-$\theta$33 & 3.2 & 29.2 & 0.052 & $33^{\circ}$ &100 & $10^8(8.6)$ & 100 & $2\times 10^4$ & 2 & 0.1 & $3.4\times 10^5$ & 90.2\\
M3.2-$\theta$63 & 3.2 & 29.2 & 0.052 & $63^{\circ}$ &100 & $10^8(8.6)$ & 100 & $2\times 10^4$ & 2 & 0.1 & $3.4\times 10^5$ & 90.2\\
\hline
M2.0-$\beta$30 & 2.0 & 10.0 & 0.027 & $13^{\circ}$ & 30 & $10^8(8.6)$ & 100 & $2\times 10^4$ & 2 & 0.1 & $3.4\times 10^5$ & 165\\
M2.0-$\beta$50 & 2.0 & 12.9 & 0.027 & $13^{\circ}$ & 50 & $10^8(8.6)$ & 100 & $2\times 10^4$ & 2 & 0.1 & $3.4\times 10^5$ & 128\\
M3.2-$\beta$30 & 3.2 & 16.0 & 0.052 & $13^{\circ}$ & 30 & $10^8(8.6)$ & 100 & $2\times 10^4$ & 2 & 0.1 & $3.4\times 10^5$ & 165\\
M3.2-$\beta$50 & 3.2 & 20.6 & 0.052 & $13^{\circ}$ & 50 & $10^8(8.6)$ & 100 & $2\times 10^4$ & 2 & 0.1 & $3.4\times 10^5$ & 128\\
\hline
M2.0-m400 & 2.0 & 18.2 & 0.013 & $13^{\circ}$ &100 & $10^8(8.6)$ & 400 & $2\times 10^4$ & 2 & 0.1 & $3.4\times 10^5$ & 22.6\\
M2.0-m800 & 2.0 & 18.2 & 0.009 & $13^{\circ}$ &100 & $10^8(8.6)$ & 800 & $2\times 10^4$ & 2 & 0.1 & $6.7\times 10^5$ & 22.3\\
M3.2-m400 & 3.2 & 29.2 & 0.026 & $13^{\circ}$ &100 & $10^8(8.6)$ & 400 & $2\times 10^4$ & 2 & 0.1 & $3.4\times 10^5$ & 22.6\\
M3.2-m800 & 3.2 & 29.2 & 0.018 & $13^{\circ}$ &100 & $10^8(8.6)$ & 800 & $2\times 10^4$ & 2 & 0.1 & $6.7\times 10^5$ & 22.3\\
\hline
M2.0-r2 & 2.0 & 18.2 & 0.027 & $13^{\circ}$ &100 & $10^8(8.6)$ & 100 & $2\times 10^4$ & 2 & 0.05 & $8.4\times 10^4$ & 22.3\\
M2.0-r0.5 &2.0 & 18.2 & 0.027 & $13^{\circ}$ &100 & $10^8(8.6)$ & 100 & $2\times 10^4$ & 2 & 0.2 & $8.4\times 10^4$ & 22.3\\
M3.2-r2 & 3.2 & 29.2 & 0.052 & $13^{\circ}$ &100 & $10^8(8.6)$ & 100 & $2\times 10^4$ & 2 & 0.05 & $8.4\times 10^4$ & 22.3\\
M3.2-r0.5 & 3.2 & 29.2 & 0.052 & $13^{\circ}$ &100 & $10^8(8.6)$ & 100 & $2\times 10^4$ & 2 & 0.2 & $8.4\times 10^4$ & 22.3\\
\hline
M2.0-2D & 2.0 & 18.2 & 0.027 & $13^{\circ}$ &100 & $10^8(8.6)$ & 100 & $2\times 10^4$ & 60 & 0.1 & $1.3\times 10^5$ & 34.6\\
M3.2-2D & 3.2 & 29.2 & 0.052 & $13^{\circ}$ &100 & $10^8(8.6)$ & 100 & $2\times 10^4$ & 60 & 0.1 & $1.3\times 10^5$ & 34.6\\
\enddata
\tablenotetext{a}{See Section \ref{sec:s2} for the convention of model name and the definition of parameters}
\tablenotetext{b}{The fiducial model.}
\vspace{-0.8cm}
\end{deluxetable*}

\begin{figure*}[t]
\vskip -0.7 cm
\hskip -0.1 cm
\centerline{\includegraphics[width=1.22\textwidth]{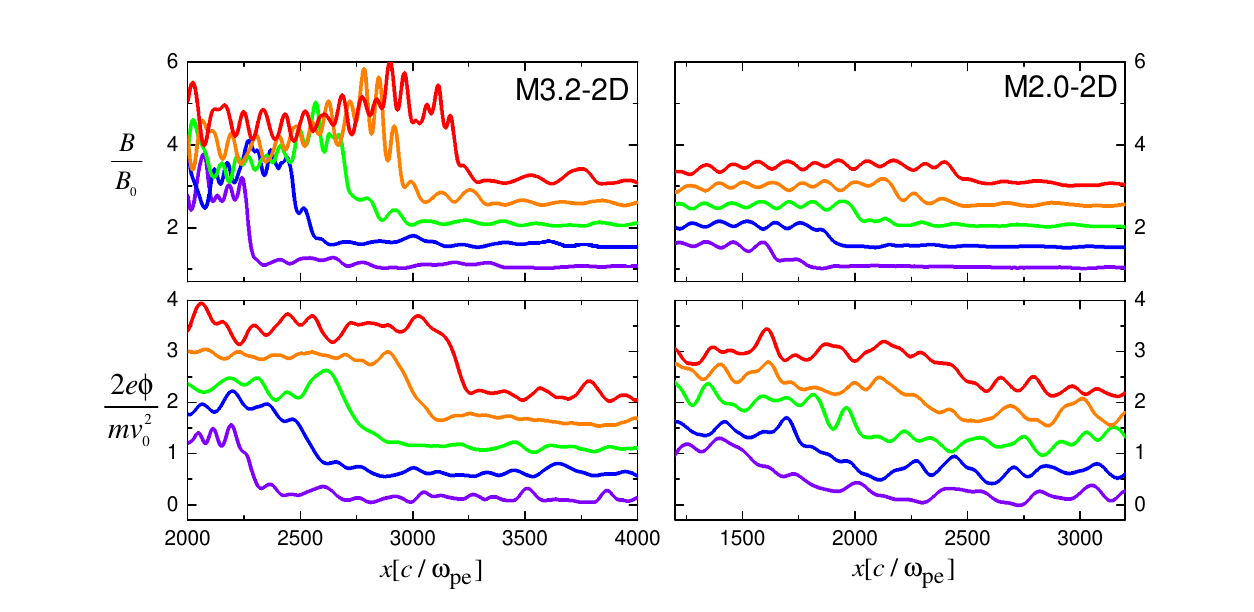}}
\vskip -0.5 cm
\caption{\label{fig:f1} Stack plots of the total magnetic field strength, $B(x)$, and the electric potential, $e\phi(x)$, averaged over the transverse direction in M3.2-2D (left) and M2.0-2D (right) models at five times from $w_{\rm pe}t = 0.8 \times 10^5$ (purple) to $ 1.2\times 10^5$ (red).}
\end{figure*}

We use an electromagnetic PIC code, TRISTAN - MP, to simulate collisionless shocks \citep{buneman1993, spitkovsky2005}. Shocks are reproduced in ``almost 1D'' (see below) or 2D planar geometry, while all the three components of particle velocity and electromagnetic fields are followed.
We adopt a simulation setup similar to those of previous works, such as that of \citet{guo2014a,guo2014b}. Magnetized plasmas with ions and electrons of Maxwell distributions move with the bulk velocity ${\mathbf{v_0}} = - v_0 \mathbf{\hat{x}}$ toward a reflecting wall at the leftmost boundary ($x = 0$), and shocks propagate along the $+\mathbf{\hat{x}}$ direction. Hence, simulations are in effect performed in the rest frame of the shock downstream flow.

In typical PIC simulations, due to severe requirements for computational resources, `ions' with reduced mass, $m_i < 1836~m_e$ ($m_i$ and $m_e$ are the ion and electron masses, respectively), are adopted to represent the real proton population.
In our simulations the ion-to-electron mass ratio, $m_i/m_e = 100-800$, is used.

The Mach number of the upstream bulk flow, $M_0$, is given as
\begin{equation}
\label{eq:e1}
M_0 \equiv \frac{v_0}{c_{\rm s}} = \frac{v_0}{\sqrt{2\Gamma k_{B}T_{i}/m_{i}}},
\end{equation}
where $c_{\rm s}$ is the sound speed in the upstream medium, $\Gamma = 5/3$ is the adiabatic index, and $k_B$ is the Boltzmann constant. 
Here, thermal equilibrium is assumed for the incoming flow, and hence the ion temperature $T_i$ is the same as the electron temperature $T_e$.
In the weakly magnetized limit (i.e., high $\beta$), the sonic Mach number, $M_{\rm s}$, of the induced shock is related to $M_0$ as
\begin{equation}
\label{eq:e2}
M_{\rm s} \equiv \frac{v_{\rm sh}}{c_{\rm s}} \approx M_0\frac{r}{r-1}.
\end{equation}
Here, $v_{\rm sh}= v_0 \cdot r/(r-1)$ is the upstream flow speed in the shock rest frame, and 
\begin{equation}
\label{eq:e3}
r = \frac{\Gamma + 1}{\Gamma - 1 + 2/M^2_{s}}
\end{equation}
is the Rankine-Hugoniot compression ratio across the shock.

\begin{figure*}[t]
\vskip -1.3 cm
\hskip -0.1 cm
\centerline{\includegraphics[width=1.08\textwidth]{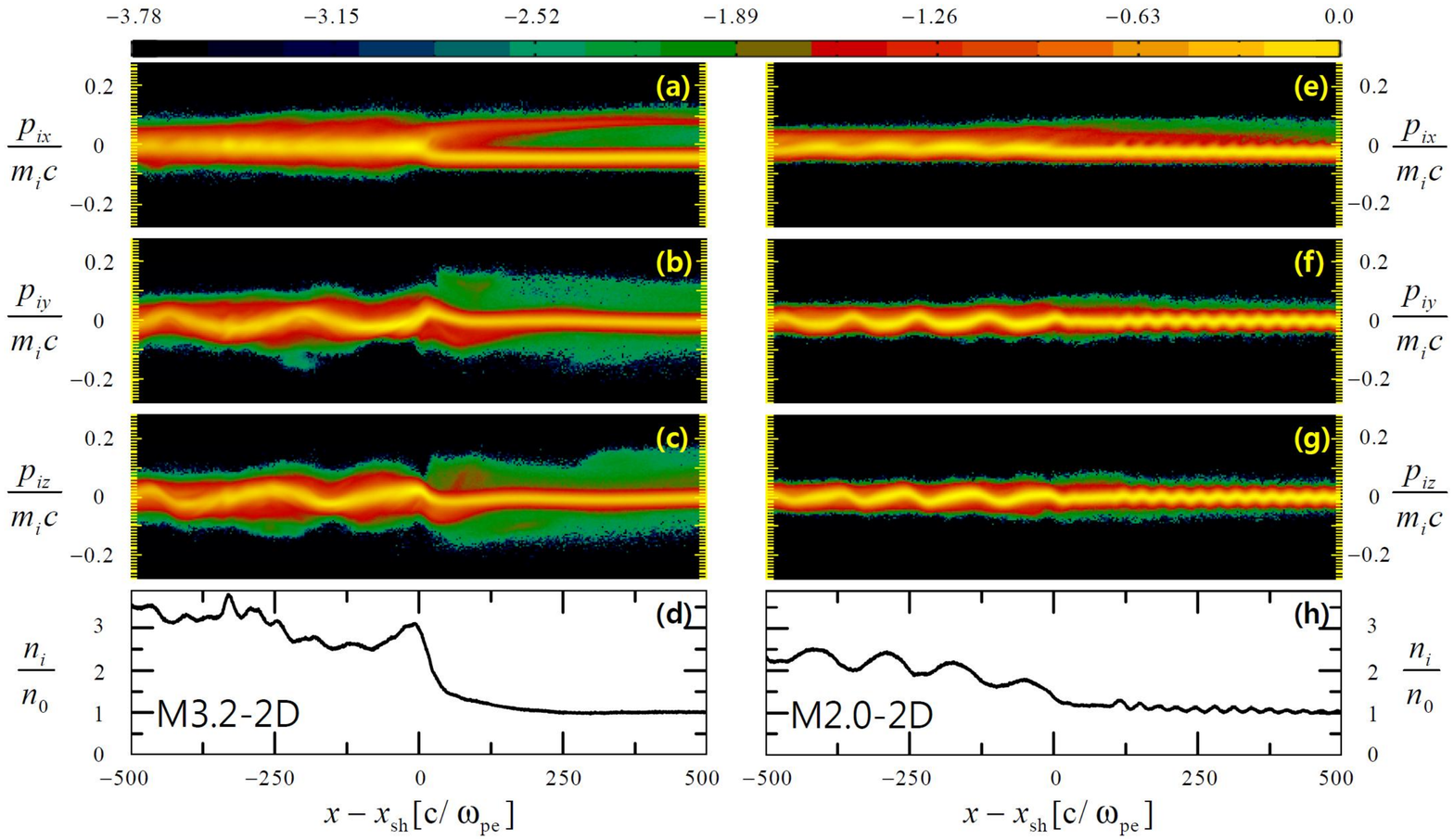}}
\vskip -1.4cm
\caption{\label{fig:f2} Shock structures in M3.2-2D (left panels) and M2-2D (right panels) models at $w_{\rm pe}t \approx 4.5\times 10^4$ ($\Omega_{\rm ci}t \approx 12$). The $x$ coordinate is measured relative to the shock position $x_{sh}$. From top to bottom, the ion distributions in the phase spaces of $p_{ix}-x$, $p_{iy}-x$, and $p_{iz}-x$, and the ion density profile are shown. The momentum is normalized by $m_{\rm i}c$, and the density is normalized by the far upstream density. The color shows the log of the ion phase-space density.}
\end{figure*}

A uniform, background magnetic field in the $x$-$y$ plane, ${\mathbf{B}_0}$, is imposed.
The strength of ${\mathbf{B}_0}$ is parameterized by $\beta$ as
\begin{equation}
\label{eq:e4}
\beta = \frac{8\pi nk_B(T_i + T_e)}{B_0^2} = \frac{2}{\Gamma}\frac{M_{\rm A}^2}{M_{\rm s}^2}.
\end{equation}
The orientation of the background magnetic field is described by the obliquity angle $\theta_{\rm Bn}$; the background magnetic field can be expressed as $\mathbf{B}_0 = B_0(\cos\theta_{\rm Bn} \mathbf{\hat{x}} + \sin\theta_{\rm Bn} \mathbf{\hat{y}})$.
\citet{caprioli2014a} showed that ion injection and acceleration in quasi-parallel shocks depends only weakly on the obliquity angle.
So we choose $\theta_{\rm Bn}=13^{\circ}$ as the fiducial value.

The shock Alfv\'en Mach number is given as
\begin{equation}
\label{eq:e5}
M_{\rm A} \equiv \frac{v_{\rm sh}}{v_{\rm A}} = \frac{v_{0}}{B_0/\sqrt{4\pi nm_i}}\frac{r}{r-1} = M_{\rm A,0}\frac{r}{r-1},
\end{equation}
where $n = n_i = n_e$ is the number density of ions and electrons in the incoming plasma,
$v_{\rm A}= B_0/\sqrt{4\pi nm_i}$ is the Alfv\'en speed along the background magnetic field,
and $M_{\rm A,0}=v_0/v_{\rm A}$ is the Alfv\'en Mach number of the upstream flow. 
The shock fast Mach number is given as $M_{\rm f}\equiv v_{\rm sh} / v_{\rm f}$, where $v_{\rm f}$ is the fast mode speed which depends on both $B_0$ and $\theta_{\rm Bn}$. For $\beta \gg 1$, $M_{\rm f}\approx M_{\rm s}$, since $v_{\rm f} \approx c_{\rm s}$.

The initial electric field is zero in our simulations. However, the incoming plasmas carry ${\mathbf{B}_0}$, and hence the motional electric field, $\mathbf{E_0} = -\mathbf{v_0}/c \times \mathbf{B}_0$, is induced, where $c$ is the speed of light.

PIC simulations follow kinetic processes on different length and time scales for different species. The electron and ion plasma frequencies are $w_{\rm pe} = \sqrt{4\pi e^{2}n/m_e}$ and $w_{\rm pi} = \sqrt{4\pi e^{2}n/m_i}$, respectively, which differ by a factor of $\sqrt{m_i/m_e}$.
We present simulation results mainly in units of $c/ w_{\rm pe}$ and $w_{\rm pe}^{-1}$, that is, the electron skin depth and the plasma oscillation period.
On the other hand, shock structures vary and evolve on the scales of the ion Larmor radius,
\begin{equation}
\label{eq:e8}
r_{\rm L,i} \equiv \frac{m_i v_0 c}{eB_0} = M_{\rm A,0}\sqrt{\frac{m_i}{m_e}}\frac{c}{w_{\rm pe}},
\end{equation}
and the ion gyration period,
\begin{equation}
\Omega_{\rm ci}^{-1} = \frac{m_i c}{eB_0} = \frac{r_{\rm L,i}}{v_0}.
\end{equation}
Hence, we interpret results in units of these scales, when necessary.

\begin{figure*}[t]
\vskip -0.95 cm
\hskip 0.2 cm
\centerline{\includegraphics[width=1.18\textwidth]{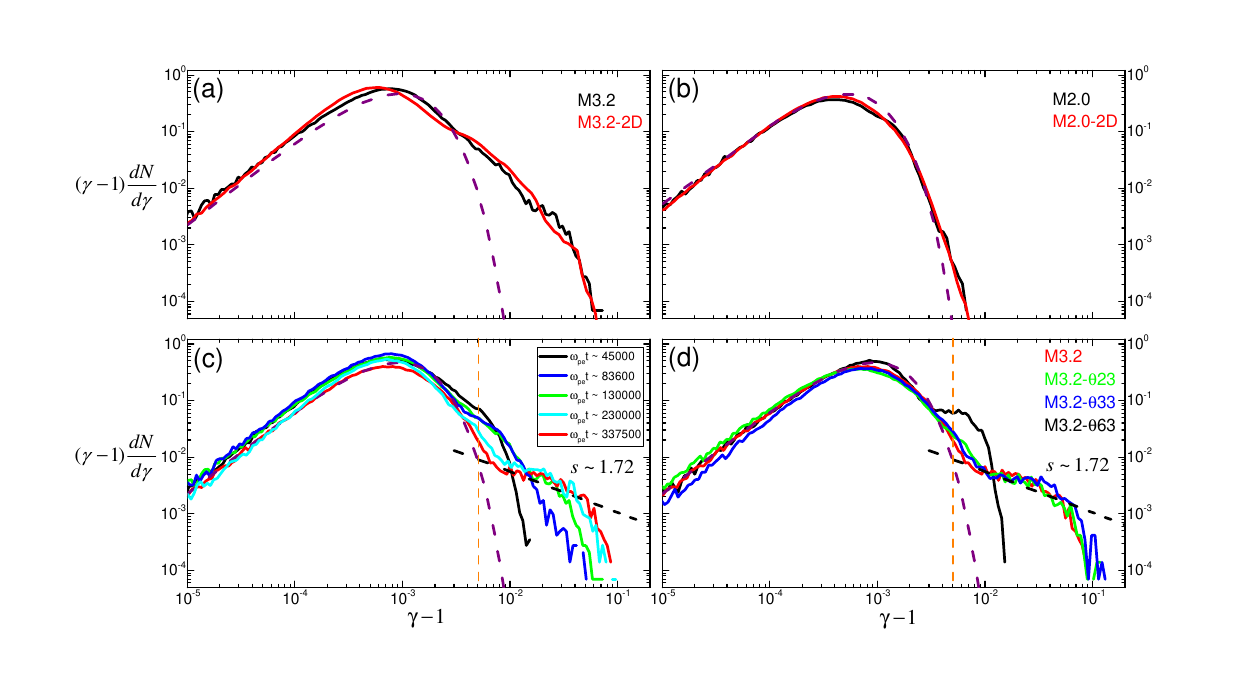}}
\vskip -1.1 cm
\caption{\label{fig:f3} 
(a) Downstream ion energy spectra at $w_{\rm pe}t \approx 1.3\times 10^5$ ($\Omega_{\rm ci}t \approx 35$) in M3.2-2D (red) and M3.2 (black) models.
(b) Downstream ion energy spectra at $w_{\rm pe}t \approx 1.3\times 10^5$ ($\Omega_{\rm ci}t \approx 35$) in M2.0-2D (red) and M2.0 (black) models.
(c) Time evolution of the downstream ion energy spectrum of M3.2 model from $w_{\rm pe}t \approx 4.5\times 10^4$ ($\Omega_{\rm ci}t \approx 12$) to $w_{\rm pe}t \approx 3.4\times 10^5 $ ($\Omega_{\rm ci}t \approx 90$). 
(d) Downstream ion spectra at $w_{\rm pe}t \approx 3.4\times 10^5$ for $M_{\rm s} \approx 3.2$ shocks with four different shock obliquity angles.
The spectra are taken from the region of $(1.5 - 2.5) r_{L,i}$ behind the shock.
The purple dashed lines show the downstream thermal Maxwellian distributions. 
In (c) and (d), the black dashed lines draw fits to nonthermal populations with the test-particle power-law slope in Equation (\ref{eq:e00}), $s \sim 1.72$, and
the orange vertical dashed lines mark the injection energy, $E_{\rm inj}\approx 5\times 10^{-3} m_i c^2$, for M3.2 model.}
\end{figure*}

\begin{figure*}[t]
\vskip -0.95cm
\hskip 0.2 cm
\centerline{\includegraphics[width=1.17\textwidth]{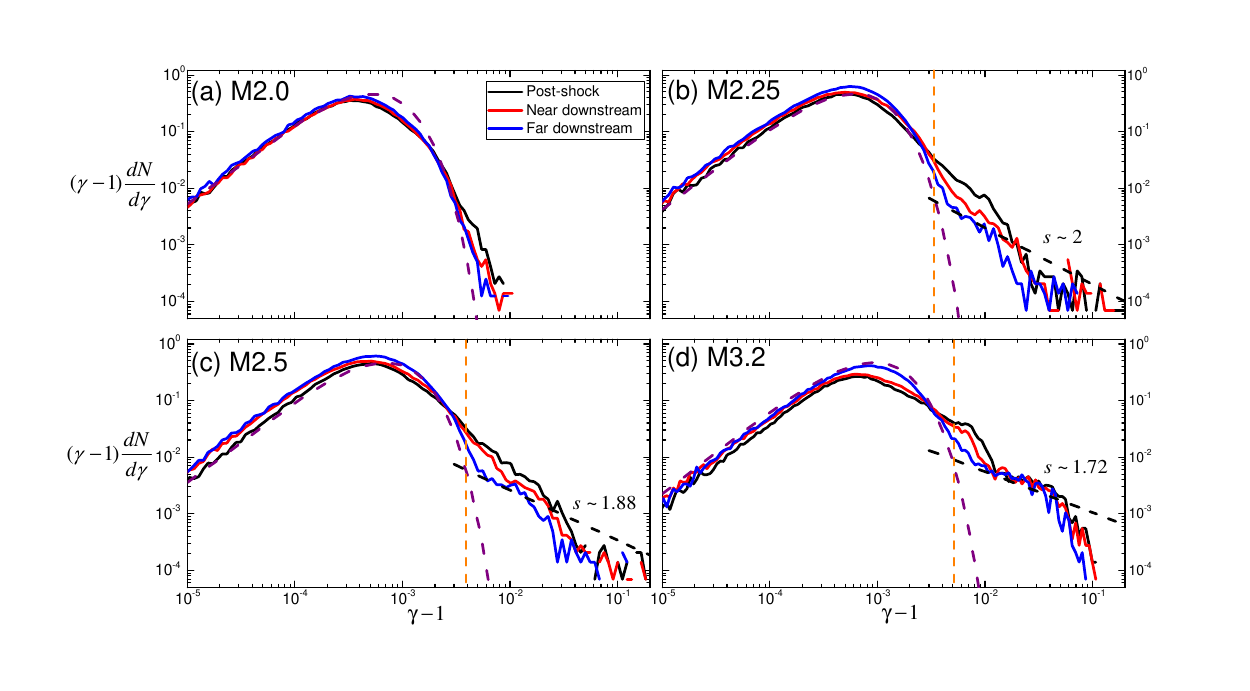}}
\vskip -1.1cm
\caption{\label{fig:f4} Downstream ion energy spectra at $w_{\rm pe}t \approx 3.4\times 10^5$ ($\Omega_{\rm ci}t \approx 90$) for M2.0, M2.25, M2.5, and M3.2 models. 
The post-shock or immediate downstream (black), near downstream (red), and far downstream (blue) energy spectra are taken from the regions of $(0 - 1) r_{L,i}$, $(1 - 2) r_{L,i}$ and $(5 - 6) r_{L,i}$, respectively, behind the shock.
The purple dashed lines show the downstream thermal Maxwellian distributions. 
In (b), (c), and (d), the black dashed lines draw fits to nonthermal populations with the test-particle power-law slopes in Equation (\ref{eq:e00}) for shocks with given $M_{\rm s}$, and the orange vertical line marks the injection energy $E_{\rm inj}$ for each model.}
\end{figure*}

We performed simulations in two-dimensional computational domains.
The longitudinal dimension, $L_x$, stretches to $2\times 10^4~c/w_{\rm pe}$. It is represented by $N_x= 2 \times 10^5$ cells
with a grid resolution of $\Delta x =0.1~c/w_{\rm pe}$ in fiducial cases. 
The transverse dimension, $L_y$, comes in two different modes: $N_y=20$ cells for ``almost 1D" simulations and $N_y=600$ cells for 2D simulations, and $\Delta y = \Delta x$ in both cases.
In each cell, 32 particles (16 per species) are placed.
The time step is $\Delta t = 0.045/w_{\rm pe}$.
From previous studies of one-dimensional PIC simulations for strong shocks \citep[e.g.,][]{park2015}, we expect that ``almost 1D" simulations would be good enough to investigate ion injection in weak ICM shocks (see Section \ref{sec:s3.1} for the dependence of simulation results on the transverse box size). 
We also expect that our results are not sensitive to the spatial resolution in our set-up (see Section \ref{sec:s3.4} for the dependence on resolution). 

The model parameters of our simulations are summarized in Table \ref{tab:t1}. 
We consider $\beta=30-100 $ and $k_BT = k_BT_e = k_BT_i = 0.0168~m_ec^2 = 8.6$~keV, relevant for typical ICM plasmas \citep[e.g.,][]{ryu2008,porter2015}. 
For given $\beta$ and $c_{\rm s}$, the incident flow velocity, $v_0$, is specified to induce shocks 
with sonic Mach number, $M_{\rm s} \sim 2 - 4$, which is characteristic for cluster merger shocks \citep[e.g.,][]{ha2018}.
M3.2 in the first row of Table \ref{tab:t1} represents the {\it fiducial} model in 1D with the following parameters: $M_{\rm s}=3.2$, $\theta_{\rm Bn}=13^{\circ}$, $\beta=100$, and $m_i/m_e=100$.
Models with different $M_{\rm s}$ are named with the combination of the letter `M' and the sonic Mach number; for example, M2.25 model has $M_{\rm s}=2.25$.
Models with parameters different from those of the fiducial model have names that are appended by a character for the specific parameter and its value.
For example, M3.2-$\theta$33 model has $\theta_{\rm Bn}=33^{\circ}$, while M3.2-m400 model has $m_i/m_e=400$.
M3.2-2D and M2-2D refer 2D models with the larger transverse dimension.
M3.2-r2 and M3.2-r0.5 models have different spatial resolutions.

The last two columns of Table \ref{tab:t1} show the end time of simulations in units of $w_{\rm pe}^{-1}$ and $\Omega_{\rm ci}^{-1}$.
For the fiducial model M3.2, $t_{\rm end} w_{\rm pe}\approx 3.4\times 10^5$, which corresponds to $t_{\rm end} \Omega_{\rm ci} \approx 90$.
The ratio of the ion gyration period to the electron oscillation period scales as $w_{\rm pe}/\Omega_{\rm ci} \propto (m_i/m_e)\sqrt{\beta}$.
So with a larger mass ratio, a longer simulation time is required to follow the formation of shocks.
Likewise, simulations with $\beta=100$ would take $\sim 10$ times longer to reach the similar stage of ion acceleration, compared to those with $\beta=1$.

\section{Results}
\label{sec:s3}

\subsection{Shock structures and ion injection}
\label{sec:s3.1}

Supercritical quasi-parallel shocks with $M_{\rm f} \gg M_{\rm f}^*$ were shown to be nonstationary and subject to quasi-periodic reformation due to the accumulation of self-generated waves in the foreshock region, resulting in time-varying overshoots in the electric shock potential and magnetic field structures \citep[e.g.,][]{caprioli2015}. At such shocks, the specular reflection of inflowing ions is thought to induce additional dissipation and supply seed particles to Fermi I acceleration, as mentioned in Introduction.
\citet{edmiston1984} presented $M_{\rm f}^*$ for $0\le \beta \le 4$, but extrapolating the result, we expect $M_{\rm f}^* \approx 1.0-1.1$ for quasi-parallel shocks in plasmas with $\beta \gg 1$. Hence, virtually all ICM shocks could be supercritical and accelerate CR protons.
However, wave excitations and wave-particle interactions themselves can provide the shock transition with {\it anomalous dissipation}, which may suppress ion reflection. 
In fact, the ion reflection process should depend on the details of kinetic processes, such as the reforming shock potential, time-varying magnetic shock ramp structures, and turbulent wave spectrum in the shock.
In addition, the particle thermal motion can smooth out the overshoot in the shock potential.
As a consequence, the reflection of ions could be suppressed at weak ICM shocks,
resulting in a higher value of $M_{\rm f}^*$ than estimated by  \citet{edmiston1984}.

The shock potential energy is estimated to be 
\begin{equation}
\label{eq:e9}
e\Delta \phi \approx \alpha(M_{\rm s},t)\frac{ m_i v_{\rm sh}^2}{2},
\end{equation}
where $\alpha$ is a factor of $\sim 1/2$ which depends on $M_{\rm s}$ and varies with time \citep{leroy1982,burgess1984}.
According to \citet{caprioli2015}, during low states without the overshoot, most of ions have $m_i v_x^2/2 > e\Delta \phi$ and advect downstream across the shock, while in high states, a substantial fraction of ions are reflected by the overshoot in $\Delta \phi$.
This results in periodic bursts of back-streaming ions along the upstream parallel magnetic fields. 
In addition, the reflected ions excite waves and amplify the transverse magnetic fields, which in turn change the quasi-parallel background fields to locally quasi-perpendicular fields (see Section \ref{sec:s3.3}).
Then, incoming ions can also be reflected in the foreshock region by those quasi-perpendicular magnetic fields through magnetic mirror \citep[e.g.,][]{treumann2008,sundberg2016}.
Obviously, shock reformation is closely related with the quasi-periodic growth and decay of the overshoot and ensuing ion reflection.
It is expected that in lower $M_s$ shocks, $\alpha(M_{\rm s})$ is reduced and the overshot does not develop, resulting in steady `subcritical' shocks.

Figure \ref{fig:f1} compares the spatial structures of the total magnetic field strength and the electric potential and their evolutions for M3.2-2D and M2.0-2D models.
The fluctuations in $B$ are mainly due to the transverse waves ($B_y$ and $B_z$), since the parallel component remains almost the same.
M3.2-2D model displays the characteristics of supercritical shocks such as
the overshoots in $\phi(x)$ and $B(x)$, self-reforming shock jump, and turbulent waves with $\langle \delta B^2 \rangle ^{1/2}/B_0 \approx 1$.
In M2.0-2D model, on the other hand, the shock is steady and smooth with much weaker waves and it does not exhibit distinct overshoots.

Figure \ref{fig:f2} shows the phase space distributions and the density profile of ions at a high state of M3.2-2D and compare with those for M2.0-2D at the same simulation time.
The presence of reflected ions moving along the $+\mathbf{\hat{x}}$ direction in the foreshock region is evident in M3.2-2D, 
while there are very little amount of ions moving upstream in M2.0-2D.
In high states of M3.2-2D, about $\sim20~\%$ of ions are reflected, while the rest get thermalized in the shock transition zone and advect downstream. The shock reformation cycle for M3.2-2D is $\sim 5.4~\Omega_{\rm ci}^{-1}$, during which the beam of reflected ions produce a new shock ramp about $2.4 ~r_{L,i}$ ahead of the original ramp. These time and length scales are larger than those for stronger shocks, considered in \citet{caprioli2015}. The mean density in the far-downstream region increases by the Rankine-Hugoniot compression factor in Equation (\ref{eq:e3}): $r\approx 3.1$ for M3.2-2D and $r\approx 2.3$ for M2.0-2D.

Figures \ref{fig:f3}(a) and (b) compare the downstream ion energy spectra, $(\gamma-1) dN/d\gamma$ in a logarithmic bin, taken from the region of $(1.5 - 2.5) r_{L,i}$ behind the shock,
in 1D (M3.2 and M2.0) and 2D (M3.2-2D and M2.0-2D) simulations  at the same simulation time.
The spectra are almost identical, indicating that `almost' 1D simulations could be employed to study the development of the spectrum, that is, the injection and early acceleration of CR protons.
Since 2D simulations are computationally much more expensive, below we use `almost' 1D simulations to investigate the effects of wide ranges of model parameters, as listed in Table \ref{tab:t1}. 

As mentioned before, the reflected ions initially gain energy via multiple cycles of SDA and become nonthermal populations with energies sufficient to diffuse across the shock.
The test-particle theory of DSA, which can be applied to weak shocks, dictates that the nonthermal momentum distribution in the downstream region is described with a power-law form of
\begin{equation}
\label{eq:e10}
f(p) \approx f_N \left({p \over p_{\rm inj}}\right)^{-q} \exp \left[-\left({p \over p_{\rm max}}\right)^2\right],
\label{finj}
\end{equation}
where $f_N$ is the normalization factor, $q=3r/(r-1)$, $p_{\rm inj}$ is the injection momentum (see below), and $p_{\rm max}$ is the maximum momentum of CR protons that increases with the shock age \citep{drury1983, kang2010}.
In the non-relativistic regime, where the CR proton energy is related to the momentum as $E =(\gamma-1 ) m_i c^2 \approx p^2/2m_{i}$, the energy distribution function for $p \lesssim p_{\rm max}$ can be approximated as
\begin{equation}
\label{eq:e00}
4\pi p^2f(p)\frac{dp}{dE} \propto \frac{dN}{d\gamma}\propto (\gamma-1 )^{-s},
\end{equation}
where $s = (q-1)/2$.

Figure \ref{fig:f3}(c) displays the time evolution of the downstream ion energy spectrum in M3.2 model, proceeding from pre-energization to the early-stage of DSA. 
By $\Omega_{\rm ci}t_{\rm end}\approx 90$ shown in the red line, the spectrum develops roughly a power-law tail with $s\approx 1.72$, the test-particle slope expected for a $M_{\rm s}=3.2$ shock. 
The downstream spectrum changes from Maxwellian to power-law distributions at $ E_{\rm inj} \approx 5.0\times 10^{-3} m_ic^2\approx 4.9 E_{\rm th}$, where $E_{\rm th} = (3/2)k_B T_2$ and $T_2$ is the downstream temperature.
Here, $E_{\rm inj}$ is the injection energy, which corresponds to the injection momentum, $p_{\rm inj}$, to mark the boundary between the thermal and nonthermal distributions.
The energized ions that belong to `a suprathermal bridge' between the thermal Maxwellian distribution and the nonthermal power-law population are often referred as suprathermal particles, as intuitively shown in Figure 2 of \citet{caprioli2014a}.
The parameter $E_{\rm inj}$ (or $p_{\rm inj}$) should depend on both $M_{\rm s}$ and $M_{\rm A}$ as well as $\theta_{\rm Bn}$, because the ability of ions to cross the shock depends on the flow compression ratio, strength of self-generated magnetic waves, and magnetic field configuration in the shock transition zone.
The result at $t_{\rm end}$ for M3.2 indicates $ p_{\rm inj}/p_{\rm th}\approx 2.7$ (where $p_{\rm th} = \sqrt{2m_i k_B T_2}$).

The long-term evolution of the downstream ion spectrum well into the full Fermi-I regime is beyond the reach of our PIC simulations.
However, the 2D hybrid simulations of \citet{caprioli2014a}, which run up to $\Omega_{\rm ci}t = 2500$, showed that
the nonthermal power-law tail extends to increasingly higher $p_{\rm max}$ with time,
and that $p_{\rm inj}/p_{\rm th}\approx 3.0-3.5$ in strong quasi-parallel shocks in $\beta\sim 1$ plasmas.
For our M3.2 model, the injection momentum at $t_{\rm end}$, $p_{\rm inj}/ p_{\rm th} \approx 2.7 $, is smaller and the normalization factor $f_N$ is higher than the values inferred from those hybrid simulations.
Such differences may come from different dimensionalities (i.e., 1D versus 2D) and different physical models (with or without electron kinetic processes) in the two simulations as well as different shock parameters (i.e. $M_{\rm s}$ and $\beta$). 
In addition, considering the trend during $\Omega_{\rm ci}t_{\rm end}\approx 35 - 90$ (or $w_{\rm pe}t_{\rm end}\approx [1.3 - 3.4] \times 10^5$) shown in Figure \ref{fig:f3}(c), we expect that $p_{\rm inj}/p_{\rm th}$ would further increase while $f_N$ decreases with time in our simulations, as high energy particles well above $p_{\rm inj}$ undergo full Fermi I acceleration.

\begin{figure}[t]
\vskip -0.3 cm
\hskip 0.05 cm
\centerline{\includegraphics[width=0.68\textwidth]{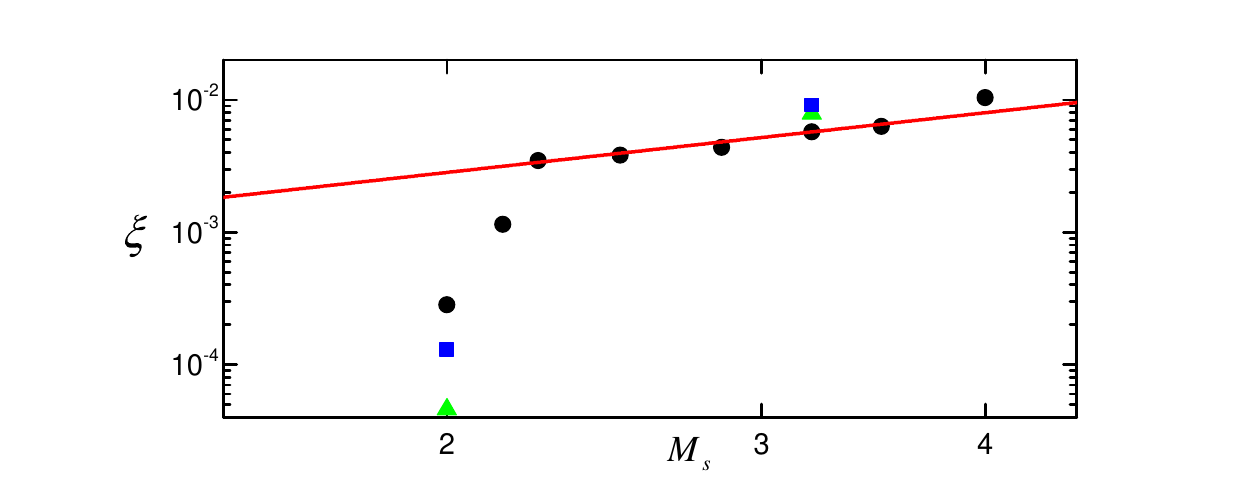}}
\vskip -0.3 cm
\caption{\label{fig:f5}Injection fraction $\xi$ at $\Omega_{\rm ci}t \approx 90$, defined in Equation (\ref{eq:e11}), for M2.0 - M4.4 models (black dots). 
The red line draws a fit, $\xi \propto M_{\rm s}^{1.5}$.
The blue squares are for M2.0-$\beta50$ and M3.2-$\beta50$ models, while the green triangles are for M2.0-$\beta30$ and M3.2-$\beta30$ models.}
\end{figure}

\begin{figure*}[t]
\vskip -0.05 cm
\hskip -0.1 cm
\centerline{\includegraphics[width=1.2\textwidth]{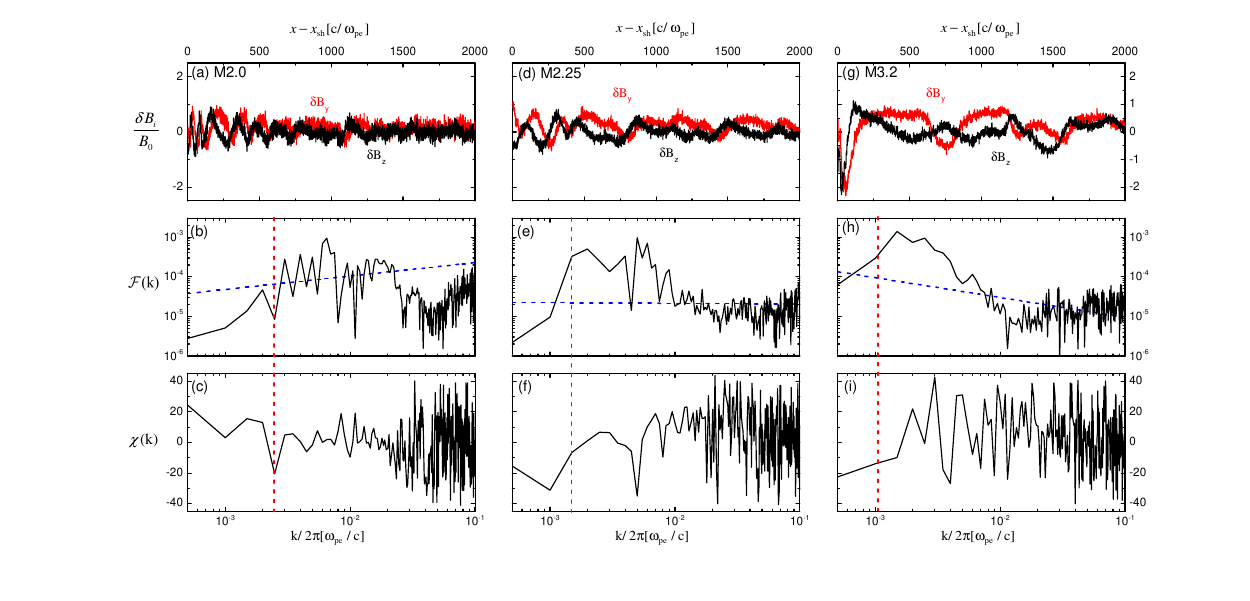}}
\vskip -0.9 cm
\caption{\label{fig:f6} Properties of self-excited magnetic fields, $\delta \mathbf{B}$, in the shock precursor ($0<(x-x_s)w_{\rm pe}/c<2 \times 10^3$) at $w_{\rm pe}t \approx 3.4\times10^5$ for M2.0, M2.25, and M3.2 models. 
Top panels: Spatial profiles of $\delta B_{y}(x)/B_0$ (Red) and $\delta B_{z}(x)/B_0$ (Black). 
Middle panels: Spectral distributions of magnetic energy, $\mathcal{F}(k)$. 
Bottom panels: Polarization angle $\chi$ in Fourier space, where $+(-)$ sign corresponds to the right-(left-)handed modes. 
The red dashed lines indicate the inverse of the mean gyroradius of nonthermal ions, while the blue dashed lines draw the characteristic power-law, $k^{q-5}$, due to resonant streaming instability.}
\end{figure*}

\begin{figure*}[t]
\vskip -0.6 cm
\hskip 0 cm
\centerline{\includegraphics[width=1.22\textwidth]{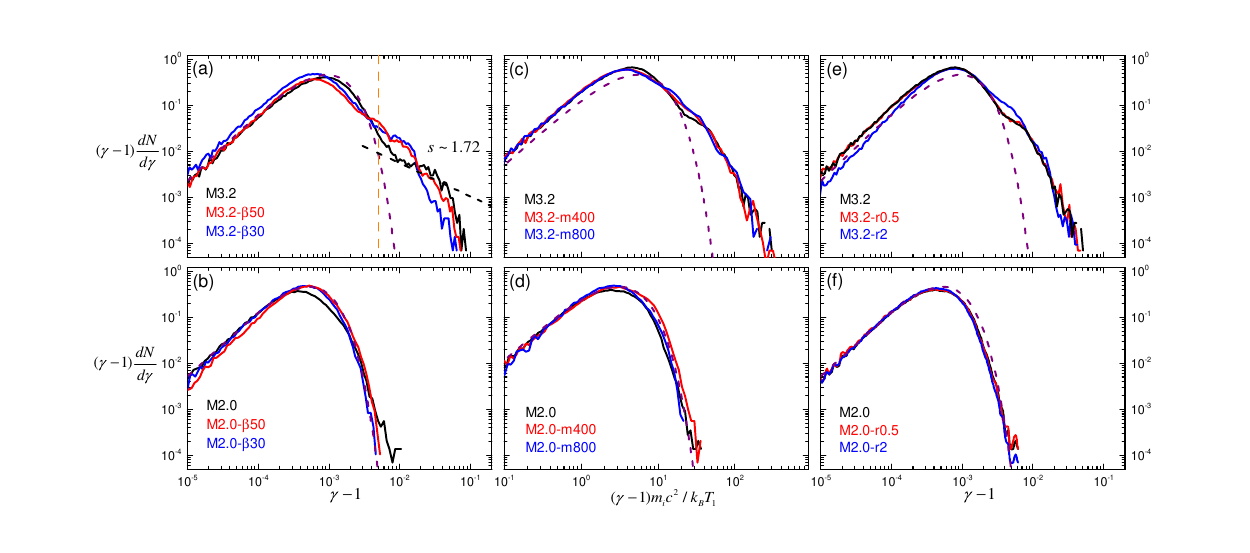}}
\vskip -0.6cm
\caption{\label{fig:f7} Downstream ion energy spectra for various M3.2 models (upper panels) and M2.0 models (lower panels). 
(a)-(b) Models with $\beta = 30$, $50$ and $100$ at $\Omega_{\rm ci}t \approx 90$ (hence, at different $w_{\rm pe}t$). (c)-(d) Models with $m_i/m_e = 100$, $400$, and $800$ at $w_{\rm pe}t \approx 8.4\times10^4$. (e)-(f) Models with different resolutions, $\Delta x = 0.2$, $0.1$, $0.05$ $c/w_{\rm pe}$, at $w_{\rm pe}t \approx 8.4\times10^4$.
The spectra are taken from the region of $(1.5 - 2.5) r_{L,i}$ behind the shock.
The purple dashed lines show the downstream thermal Maxwellian distributions.
In (a), the black dashed line draws a fit to nonthermal populations with the test-particle power-law slope in Equation (\ref{eq:e00}), $s \sim 1.72$, and
the orange vertical dashed line marks the injection energy, $E_{\rm inj}\approx 5\times 10^{-3} m_i c^2$, for $M_{\rm s} = 3.2$ models.}
\end{figure*}

Figure \ref{fig:f4} shows the ion energy spectra in three downstream regions, immediate, near, and far downstream, at $t_{\rm end}$ in 1D simulations for M2.0, M2.25, M2.5, and M3.2 models. 
Models with $M_{\rm s} \ge 2.25$ show the development of nonthermal power-law-like tails with slopes, consistent with the test-particle values for given $M_{\rm s}$.
On the other hand, the shock with $M_{\rm s} = 2$ does not possess any appreciable population of nonthermal particles beyond the Maxwellian distribution.
From these and also the spectrum for M2.15 (not shown in the figure), we estimate that in ICM plasmas with $\beta\approx 100$, the fast first critical Mach number occurs at $M_{\rm f}^* \approx 2.25$, which is higher than the value $M_{\rm f}^* \approx 1.0 - 1.1$, quoted from \citet{edmiston1984}.
In summary, our results suggest that only in ICM shocks with $M_{\rm s} \gtrsim 2.25$, a substantial fraction of incoming protons are injected into the Fermi I process, and then are expected to be accelerated to high energy CRs. 
At $M_{\rm s} < 2.25$, the DSA of CR protons may not occur.

Although our simulations extend only to very early stages of DSA, 
we attempt to estimate the `ion injection fraction' at the end of simulation time $t_{\rm end}$.
We define it as the nonthermal ion fraction in the downstream region,
\begin{equation}
\label{eq:e11}
\xi \equiv \frac{1}{n_2}\int_{p_{\rm min}}^{p_{\rm max}} 4\pi \langle f(p, t_{\rm end})\rangle p^2 dp , 
\end{equation}
where $p_{\rm max}$ is given in Equation (\ref{eq:e10}) and $\langle f(p)\rangle $ is averaged over the region of $(1.5 - 2.5) r_{L,i}$ behind the shock.
Here, $p_{\rm min}$ should be somewhat larger than $p_{\rm inj}$, and we arbitrary choose $p_{\rm min} = \sqrt{2}p_{\rm inj}$.
Figure \ref{fig:f5} shows $\xi$ for 1D models with a range of shock Mach numbers, $M_{\rm s}\approx 2.0-4.0$, and $\theta_{\rm Bn} = 13^{\circ}$.
As shown in the red line, it increases with the Mach number, roughly as $\xi(M_{\rm s}) \propto M_{\rm s}^{1.5}$ for $M_{\rm s} \gtrsim 2.25$.
In weaker shocks, the shock compression ratio is smaller and the fractional energy gain at each SDA cycle is smaller.
Hence, ions need to undergo more cycles of reflection to be injected to DSA, leading to a smaller injection fraction.
Moreover, the factor $\alpha$ in $\Delta \phi$ in Equation (\ref{eq:e9}) should decrease with decreasing $M_{\rm s}$.
As a result, the injection fraction is expected to be smaller in lower $M_{\rm s}$ shocks. Our simulations indicate that the injection faction drops rather abruptly to very small values for $M_{\rm s} < 2.25$, confirming that the fast first critical Mach number could be $M_{\rm f}^* \approx 2.25$.

In the test-particle limit of low $M_{\rm s}$ shocks, where the momentum distribution of nonthermal particles is given as in Equation (\ref{eq:e10}), $\xi$ depends mainly on the normalization factor, $f_N$, and the minimum momentum $p_{\rm min}$.
The maximum momentum, $p_{\rm max}$, which increases with time, does not affect $\xi$ much.
As mentioned above, $f_N$ seems to decrease while $p_{\rm inj}$ increases  in our simulations; then, $\xi$ also decreases with time.
Hence, we expect that in the DSA regime, $\xi$ would be smaller than that presented in Figure \ref{fig:f5}.
The `converged' $\xi$ should be estimated through simulations that extend to much larger number of ion gyration periods.
Thus, here the relative trend of $\xi$ as a function of $M_{\rm s}$ only matters.

\subsection{Magnetic field amplification}
\label{sec:s3.2}

As shown in Figures \ref{fig:f1} and \ref{fig:f2}, in supercritical quasi-parallel shocks, a substantial fraction of ions are reflected by the shock potential barrier and stream along the direction parallel to upstream magnetic fields, and then excite turbulent waves via CR-driven instabilities.
To present the properties of excited waves, we show the spatial distributions of self-generated magnetic field components, $\delta B_{y}$ and $\delta B_{z}$, and their Fourier analyses in the shock precursor. Here, $\delta B_{y} = B_{y} - B_0 \sin\theta_{\rm Bn}$ and $\delta B_{z} = B_{z}$.

Using the stationary equation for the growth and transport of magnetic turbulence \citep{mckenzie1982}, \citet{caprioli2014b} derived the Fourier-space behavior of the magnetic energy of the Alfv\'en waves produced by `resonant streaming instability' in the upstream region of strong shocks with the momentum distribution of nonthermal particles of $f(p) \propto p^{-4}$. The analysis can be straightforwardly extended to include low $M_{\rm s}$ shocks, that is, for $f(p) \propto p^{-q}$ with $q\ge 4$.
Assuming equipartition between the electromagnetic and kinetic energy densities in the waves, it can be shown that the magnetic energy per unit logarithmic bandwidth with wavenumber $k$ is given as
\begin{equation}
\label{eq:e14}
\mathcal{F}(k) \propto k \left(\frac{\delta B}{B_0}\right)^2 \propto k^{q-5},
\end{equation}
and hence $\delta B_{y,z}/B_0 \propto k^{(q-6)/2}$.
The dependence on the sonic Mach number, $M_{\rm s}$, enters through $q$.
For strong shocks with $q = 4$, $\mathcal{F}(k) \propto k^{-1}$, recovering the derivation of \citet{caprioli2014b}.

Note that if resonant modes are excited by the streaming CRs with power-law spectral slope $q > 5$, Equation (\ref{eq:e14}) predicts that most of the magnetic energy powers will be at large wavenumbers.
The Mach number that divides the positive and negative slopes of $\mathcal{F}(k)$ is $M_{\rm s} = \sqrt{5} = 2.24$, which gives $q=5$. 

The polarization angle $\chi$ of elliptically polarized waves can be used to quantify the handedness of waves \citep[e.g.,][]{park2015};
\begin{equation}
\chi (k) = \frac{1}{2}\sin^{-1}\left(\frac{V}{I}\right)
\end{equation}
is calculated as a function of $k$, where $I(k) = |B_{y}(k)|^2 + |B_{z}(k)|^2$ and $V(k)= |B_{y}(k)|^2 - |B_{z}(k)|^2$.
Positive (negative) values of $\chi$ indicate right-handed (left-handed) polarizations of waves.
In particular, $\chi= + (-) 45 ^{\circ}$ corresponds to right-handed (left-handed) circularly polarized waves.

Figure \ref{fig:f6} shows $\delta B_{y}(x)$ and $\delta B_{z}(x)$, $\mathcal{F}(k)$, and $\chi(k)$ in the region of width $\sim 2 \times 10^3~c/ w_{\rm pe}$, upstream of the shock ramp, in M2.0 ($M_{\rm A}\approx 18$), M2.25 ($M_{\rm A}\approx 20$), and M3.2 ($M_{\rm A}\approx 29$) models at the end of simulations.
Waves are present in all three models. 
The red vertical lines mark $k_{\rm CR} = 2 \pi/\rho_{\rm CR}$, where $\rho_{CR}$ is the average gyroradius of nonthermal ions in the upstream region of $(x-x_s)w_{\rm pe}/c = [1.75-2.0] \times 10^3$, for each model.
At higher $M_{\rm s}$, ions are accelerated to higher energies, and hence $\rho_{\rm CR}$ is larger. Also the characteristic wavelength of excited waves is longer, as can be seen in the top panels.
Moreover, since more ions are reflected at higher $M_{\rm s}$, the wave amplitude, represented by the magnitude of $\delta B_{y}$ and $\delta B_{z}$, is larger.

\begin{figure*}[t]
\vskip 0.25 cm
\hskip -0.15 cm
\centerline{\includegraphics[width=1.22\textwidth]{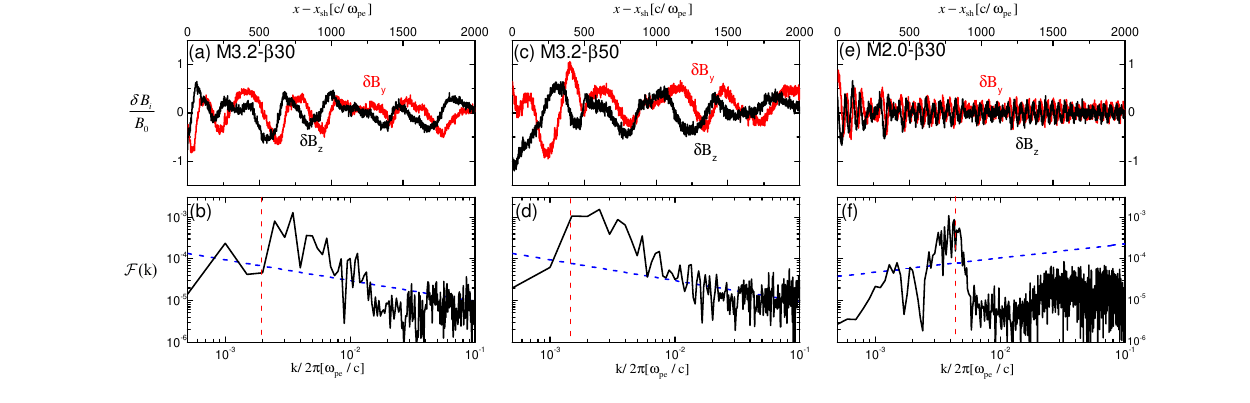}}
\vskip -0.4 cm
\caption{\label{fig:f8} Properties of self-excited magnetic fields, $\delta \mathbf{B}$, in the shock precursor ($0<(x-x_s)w_{\rm pe}/c<2 \times 10^3$) at $\Omega_{\rm ci}t \approx 90$ for M3.2-$\beta$30, M3.2-$\beta$50, and M2.0-$\beta$30 models. 
Upper panels: Spatial profiles of $\delta B_{y}(x)/B_0$ (Red) and $\delta B_{z}(x)/B_0$ (Black). 
Lower panels: Spectral distribution of magnetic energy, $\mathcal{F}(k)$. 
The red dashed lines indicate the inverse of the mean gyroradius of nonthermal ions, while the blue dashed lines draw the characteristic power-law, $k^{q-5}$, due to resonant streaming instability.}
\end{figure*}

The spectral distributions of $\mathcal{F}(k)$ (so $\delta B(k)$) and $\chi(k)$ for M3.2 model (right-hand panels) bear similarities, for instance, to those of a $M_{\rm A}=20$ ($M_{\rm s}=40$) shock presented in \citet{park2015}. In both cases, a substantial fraction of ions are injected, although shock parameters are different.
As noted in the Introduction, \citet{caprioli2014b} argued that $M_{\rm A} \sim 30$ roughly divides the Mach number ranges where resonant streaming instability and nonresonant current driven instability dominantly operate.
M3.2 model has $M_{\rm A}\approx 29$, close to the boundary value.
The figures for M3.2 show that in the range of $k \gtrsim k_{\rm CR}$, modes with both positive and negative $\chi$ are present, but those with positive $\chi$ are more frequent.
The spectrum, $\mathcal{F}(k)$, shows a distribution steeper than the quasilinear turbulence spectrum, $k^{q-5}$ (blue dashed line), for the waves excited by resonant streaming instability.
These indicate that nonresonant modes are likely dominant in the range of $k \gtrsim k_{\rm CR}$.
For $k \lesssim k_{\rm CR}$, however, $\chi$ is negative, and hence resonant modes operate.
M2.25 model in the middle panels, in which the reflection and injection of ions are still observed, shows similar trends, although the injection fraction is smaller. Especially, for $k \lesssim k_{\rm CR}$, $\chi$ is negative, indicating the operation of resonant modes.

On the other hand, M2.0 model in the left-hand panels shows different behaviors. 
Through hybrid simulations of weak shocks in $\beta \sim 1$ plasmas, \citet{kraussvarban1991} showed that fast magnetosonic waves with large-amplitudes $\delta B/B_0 \sim 1$ and characteristic wavenumbers $ck/w_{\rm pi} \lesssim 1$ (corresponding to $ck/(2\pi w_{\rm pe}) \lesssim 0.02$ with $m_i/m_e=100$ in our PIC simulations) are present in the shock upstream region. 
Hence, the modes with large amplitudes of $\mathcal{F}(k)$ in the range of $2\times 10^{-3} \lesssim ck/(2\pi w_{\rm pe}) \lesssim 3\times 10^{-2}$ in Figure \ref{fig:f6}(b)  are probably the same kind of waves excited by a very small amount of reflected ions escaping to the upstream region, as can be seen in Figure \ref{fig:f2}.
They are likely to be the right-handed whistler waves generated by the resonant ion/ion beam instability, which do not scatter ions resonantly. 
This could be the reason why reflected ions are not efficiently accelerated to suprathermal energies and the injection fraction, $\xi$, abruptly drops in shocks with $M_{\rm s} < 2.25$ (see Figure \ref{fig:f5}).

\subsection{Dependence on shock parameters}
\label{sec:s3.3}

In this section, we investigate the dependence of ion injection on shock obliquity angle and plasma beta.
Figure \ref{fig:f3}(d) compares models with different $\theta_{\rm Bn}$ for $M_{\rm s}=3.2$ shocks.
It is known that ion injection and acceleration depend only weakly on $\theta_{\rm Bn}$ for quasi-parallel shocks \citep{caprioli2014a}, and M3.2 (with $\theta = 13^{\circ}$), M3.2-$\theta$23, and M3.2-$\theta$33 models show consistent results.
With higher obliquity angles, the injection energy, $E_{\rm inj}$, would be higher, and so a larger number of SDA cycles would be required for injection to DSA \citep{caprioli2015}. 
However, the ion injection fraction $\xi$ for quasi-parallel shocks, estimated from the figure, shows only a weak dependence on the shock obliquity as long as $\theta_{\rm Bn}\lesssim 45^{\circ}$.
On the other hand, ion injection is expected to be severely suppressed for quasi-perpendicular shocks \citep{caprioli2014a}, and the results for M3.2-$\theta$63 model confirm it. Despite that ions are reflected more efficiently and a significant fraction ($\sim 20 \%$) of downstream ions form a suprathermal component in quasi-perpendicular shocks, the reflected ions do not gain sufficient energies for injection to the full Fermi I process before they advect downstream behind the shock.

Figure \ref{fig:f7}(a) presents the dependence on $\beta$ for $M_{\rm s} = 3.2$ shocks at the same ion gyration periods.
Except $\beta$, other shock parameters, i.e., $\theta_{\rm Bn}$, $n$, and $T$, are fixed. Models of smaller $\beta$ have smaller $M_{\rm A}$: $M_{\rm A} \approx 16,$ $21$, and $29$ for $\beta = 30$, $50$, and $100$, respectively.
With gyroradius $\propto B^{-1}$, the resonant wavelengths of upstream nonthermal ions roughly scale as $k_{\rm CR} \propto \beta^{-1/2}$.
They are drawn with the red dashed lines in the lower panels of Figure \ref{fig:f8}.
The characteristic wavelength of resonantly excited waves is also shorter for smaller $\beta$, as shown in the upper panels of Figures \ref{fig:f6} and \ref{fig:f8}.
\citet{caprioli2014b} argued that $(\delta B/ B_0 )^2 \propto M_{\rm A} \propto \beta^{1/2}$, and hence turbulence is weaker for smaller $\beta$. 
The panels for $\mathcal{F}(k)$ in Figures \ref{fig:f6} and \ref{fig:f8} confirm that the amplitude of turbulent magnetic field spectrum is smaller for smaller $\beta$. 
With weaker levels of magnetic turbulence, the development of nonthermal population is less efficient. 
The ion spectra for $\beta = 50$ and $30$ in Figure \ref{fig:f7}(a) look similar to those of earlier epochs for $\beta = 100$ case (see Figure \ref{fig:f3}(c)).

The injection fractions, $\xi$, for $M_{\rm s} = 3.2$ shocks with $\beta=50$ and 30 are a slightly larger than that for $\beta=100$, as shown by the blue square and green triangles in Figure \ref{fig:f5}, respectively.
This is consistent with the fact that the downstream ion spectra develop more slowly at lower $\beta$, while $\xi$ decreases with time in our simulations.
The dependence of $\xi$ on $\beta$, hence, should be investigated with simulations that extend much longer,
although we expect that $\xi$ would be insensitive to $\beta$ in the DSA regime.

Figure \ref{fig:f7}(b) presents the downstream ion energy spectra for $M_{\rm s}=2.0$ shocks with different $\beta$. The reflection of ions are inefficient, regardless of $\beta$, at such low $M_{\rm s}$ shocks. Accordingly, the injection fraction is small, as shown in Figure \ref{fig:f5}. Hence, as remarked in Section \ref{sec:s3.1}, we expect that the DSA of CR protons would be inefficient in weak shocks, for instance, those with $M_{\rm s} < 2.25$, in ICM plasmas if their $\beta$ is high in the range of $30-100$.

\subsection{Dependence on simulation parameters}
\label{sec:s3.4}

We here examine how our findings depend on simulation parameters, that is, the ion-to-electron mass ratio, $m_i/m_e$, and the spatial resolution, $\Delta x$.
We use the reduced ion mass of $m_i=100~m_e$ as the fiducial value. Hence, the ion thermal speed is higher by a factor of $\sqrt{18.36}$ than in reality for a given preshock temperature. The ion gyroradius is reduced by the same factor, which results in smaller differences in ion and electron penetration-depths at the shock transition. However, the potential energy drop at the shock, $e\Delta \phi \propto m_i v_0^2$, is independent of $m_i/m_e$, and hence the reflection capability of the shock is expected to be insensitive to it.
Figures \ref{fig:f7}(c) and (d) demonstrate that our simulation results do not sensitively depend on the mass ratio. Hence, we expect that the critical Mach, $M_{\rm s}^*\approx 2.25$, remains the same even for the cases with more realistic mass ratio.

Figures \ref{fig:f7}(e) and (f) explore the dependence of the ion energy spectra on the grid resolution.
Our simulations with different resolutions produce essentially the same ion spectra, especially for the nonthermal population.

\section{Summary}
\label{sec:s4}

In supercrical quasi-parallel shocks, a substantial fraction of incoming ions are specularly reflected by the overshoot in the electrostatic shock potential and locally perpendicular magnetic fields in the shock transition \citep{treumann2009}.
The reflected ions escaping upstream along parallel magnetic fields excite left-handed (right-handed) polarized waves via CR-driven resonant (nonresonant) instabilities, amplifying the transverse components of magnetic fields \citep{caprioli2014b}. This induces again locally quasi-perpendicular fields in the shock ramp, which in turn facilitate the specular reflection of ions due to magnetic deflection and the SDA through the drift along the motional electric field \citep{sundberg2016}. 
In $\beta\sim 1$ environments, the critical fast Mach number was predicted to be $M_{\rm f}^* \approx 1.0-1.5$ for shock obliquity angle $\theta_{\rm Bn} \lesssim 45^{\circ}$ from a fluid approach \citep{edmiston1984}.
In this paper, we study kinetic plasma processes and estimate the critical sonic Mach number $M_{\rm s}^*$, above which ion reflection and acceleration is efficient,
by performing PIC simulations for quasi-parallel shocks in ICM plasmas with high $\beta$.
A wide range of shock parameters are considered, as listed in Table \ref{tab:t1}.

The main results are summarized as follows:

1. We find that the critical Mach number is $M_{\rm s}^*\approx 2.25$ for quasi-parallel shocks in high $\beta$ environments, which is higher than the value
$M_{\rm f}^*\approx 1.0-1.1$ estimated from the MHD Rankine-Hugoniot relation by \citet{edmiston1984}.
We conjecture that the anomalous dissipation inside the shock transition due to wave-generations may provide the necessary shock dissipation in shocks with $M_{\rm s} < 2.25$, 
since higher $\beta$ plasmas are more prone to the self excitation of waves such as dispersive magnetosonic whistlers.
Only in supercritical quasi-parallel shocks with $M_{\rm s} \gtrsim 2.25$, a substantial fraction of ions impinging on the shock ramp are reflected to upstream and gain sufficient energies via SDA to become nonthermal particles.

2. In ICM plasmas with $\beta \approx 100$, even weak shocks with $M_{\rm s}\approx 2-4$ 
have relatively large Alfv\'en Mach numbers, $M_{\rm A}\approx \sqrt{\beta} M_{\rm s}\approx 20-40$.
With $(\delta B/ B_0)^2 \propto M_{\rm A}$ for Alfv\'en waves generated via CR-ion driven instabilities, magnetic turbulence should be strong even in such weak shocks.
According to \citet{caprioli2014b}, while resonant instability is dominant for $M_{\rm A} \lesssim 30$, nonresonant instability grows faster for $M_{\rm A} \gtrsim 30$.
In our fiducial model M3.2 with $M_{\rm s}\approx 3.2$ and $M_{\rm A}\approx 29$, hence, both instabilities are expected to operate.
We find that in the range of $k \gtrsim k_{\rm CR}$ (where $k_{\rm CR}$ is the wavenumber corresponding to the average gyroradius of nonthermal ions in the upstream region), nonresonant modes are likely dominant.
In the range of $k \lesssim k_{\rm CR}$, on the other hand, mostly resonant modes appear.

3. We estimate the CR injection fraction, $\xi$, defined as the number fraction of nonthermal ions with $p \geq \sqrt{2} p_{\rm inj}$ (so $E \gtrsim 10 E_{\rm th}$) in the downstream spectrum, as a function of $M_{\rm s}$, at the end of 1D simulations, which corresponds to the very early stage of DSA.
It ranges $\xi \sim 10^{-3} - 10^{-2}$ for quasi-parallel shocks with $M_{\rm s} = 2.25 - 4.0$, and increases with sonic Mach number as $\xi \propto M_{\rm s}^{1.5}$.
Below $M_{\rm s} = 2.25$, $\xi$ drops sharply, indicating inefficient injection of ions.
Our simulations indicate that $\xi$ decreases with time, and hence the values in the full Fermi I regime would become smaller.
We estimate $\xi$ mainly for nearly parallel shocks with $\theta_{\rm Bn}=13^{\circ}$ in $\beta \approx 100$ plasmas. The fraction, $\xi$, is expected to depend only weakly on the shock obliquity as long as $\theta_{\rm Bn}\lesssim 45^{\circ}$, while $\xi$ seems to be slightly larger for smaller $\beta$ in the range of $\beta = 30 - 100$. 
The estimate of $\xi$ for broad ranges of $\theta_{\rm Bn}$ and $\beta$ as well as $M_{\rm s}$ requires a larger set of simulations.
Especially, the estimate of `converged' $\xi$ in the DSA regime requires simulations extending to very large numbers of ion gyration periods, perhaps simulations with tools other than PIC codes. Since it is beyond the scope of this study, and we leave it as a future work.

4. If quasi-parallel ICM shocks with $M_{\rm s} < 2.25$ were unable to generate CR protons as implied by this study, the level of gamma-ray flux due to shock-accelerated CR protons would be much lower than previously estimated \citep[e.g.,][]{vazza2016}. This may provide a clue to the mystery of non-detection of gamma-ray emission from galaxy clusters \citep{ackermann2016}.

\acknowledgments
The work was supported by the National Research Foundation (NRF) of Korea through grants 2016R1A5A1013277, 2017R1A2A1A05071429, and 2017R1D1A1A09000567. J.-H.H. was supported by the Global PhD Fellowship of the NRF through 2017H1A2A1042370.

\end{document}